\documentclass[12pt]{article}
\usepackage{amsmath,amssymb,graphicx,hyperref,natbib}
\usepackage[a4paper,margin=1in]{geometry}
\usepackage{setspace}
\usepackage{titlesec}
\usepackage{lmodern}
\usepackage[utf8]{inputenc}
\usepackage{textgreek}
\title{\textbf{Institutional Noise, Strategic Deviation, and Intertemporal Collapse:\\ A Formal Model of Miner Behaviour under Protocol Uncertainty}}
\author{Craig S. Wright\\
\textit{University of Exeter, Business School}\\
\texttt{cw881@exeter.ac.uk}}
\date{}

\begin{document}
\maketitle

\begin{abstract}
This paper introduces a formalised economic model of miner behaviour in blockchain systems where institutional rules—i.e., protocol-level parameters—are mutable. By integrating Austrian time preference theory with dynamic game theory, we simulate the strategic calculus of rational actors under stochastic institutional noise. The model demonstrates that when protocol predictability degrades, the equilibrium structure of miner engagement collapses. Agents rationally defect, collapse cooperative equilibria, and engage in rent-seeking or exploitative meta-strategies. Simulation results show a critical threshold beyond which long-term investment in computational capital becomes economically irrational. These findings support the view that institutional immutability is not merely normative but structurally necessary for long-horizon sustainability in blockchain networks.
\end{abstract}

\noindent\textbf{Keywords:} time preference, institutional economics, game theory, blockchain governance, protocol mutability, miner strategy, Austrian capital theory, calculability, dynamic equilibria, rent-seeking, protocol noise, entrepreneurial planning, strategic deviation

\section{Introduction}

Incentives operate not in a vacuum but within structured rule environments, and when those rules are mutable, the very calculability of economic action is undermined. This paper builds a formal model linking institutional mutability—specifically in blockchain protocols—to the breakdown of cooperative equilibria among economically rational miners. The analysis is rooted in Austrian capital theory, with its emphasis on roundabout production and time preference, and in game theory, where repeated strategic interaction underpins stability. When protocols change unpredictably, agents adapt not by reinforcing productive engagement but by shifting to meta-strategic behaviours—lobbying, short-term arbitrage, or hedging against rule shocks. This introduces discounting distortions similar to those seen in fiat monetary regimes, collapsing the long-horizon calculability that undergirds entrepreneurial investment. By extending the theoretical structure of Wright (2025), we simulate payoff transformations under stochastic protocol environments, demonstrating threshold effects and collapse zones in cooperative engagement. This work bridges game theory, Austrian economics, and institutional cryptoeconomics to argue that immutability is not merely desirable but necessary for sustainable incentive coherence. The implications extend beyond Bitcoin, pointing to foundational principles for digital governance: rules must be fixed not only to coordinate behaviour, but to make calculation itself possible.

\subsection{Overview and Context}

This paper presents a formal analysis of miner strategy formation under conditions of protocol rule mutability, integrating Austrian capital theory and repeated game models into a unified treatment of institutional cryptoeconomics. The core thesis is that unpredictability in protocol governance generates distortions analogous to monetary instability, leading to the breakdown of cooperative equilibria. Building on Wright (2025), the model traces how changes to rule structures—such as block size limits, relay policy, and mempool behaviour—alter strategic payoff matrices, induce higher discount rates, and catalyse deviant miner conduct. The context is both empirical and theoretical: recent events in BTC Core reveal how small, coordinated groups have incrementally centralised power by layering discretionary rule adjustments under the guise of protocol maintenance. By contrast, the original Bitcoin protocol offered a fixed rule-set that enabled long-term planning, calculability, and economic coordination. This study situates that contrast within a formal structure, demonstrating how mutable systems encourage non-productive rent-seeking, while fixed systems foster sustainable cooperation. The sections that follow articulate the philosophical and theoretical grounding, construct the dynamic game-theoretic framework, simulate payoff structures under rule shock volatility, and derive broader implications for decentralised infrastructure.

\subsection{Economic Significance of Institutional Stability}

Institutional stability serves as the precondition for rational economic calculation, a principle deeply embedded in the Austrian tradition. Without predictable rules, entrepreneurs cannot form expectations about future states, rendering capital allocation speculative rather than strategic. Mises (1949) argued that calculability is only possible within a framework of stable institutional constraints, a position echoed by Hayek in his exposition on spontaneous order. In blockchain economies, the protocol assumes this constitutional role. It sets the boundaries within which agents operate, form strategies, and undertake long-term investments. When these rules are mutable, however, the structure of expectations disintegrates: the future becomes opaque, time preferences rise, and actors optimise for short-term gain rather than durable contribution. Empirical deviations observed in BTC Core—such as the rise of empty blocks, manipulation of relay policies, and off-chain centralisation—are not incidental but symptomatic of deeper institutional volatility. The game becomes a hypergame, where players not only choose actions but attempt to redefine the rules. As Wright (2025) formalised, even marginal uncertainty in rule persistence translates into elevated discounting and undermines the viability of cooperative strategies. Stability is not an aesthetic preference—it is an economic necessity. Without it, market coordination collapses, and the infrastructure reverts from a system of order to a battlefield of political arbitration.

\subsection{Model Scope and Theoretical Integration}

The model developed herein integrates Austrian economics and dynamic game theory to analyse strategic miner behaviour under differing institutional regimes—specifically, fixed versus mutable protocol rule-sets. The scope of the model is both narrow and expansive: narrowly, it focuses on rational miner incentives in a repeated game environment where protocol conditions evolve; expansively, it offers a general framework for interpreting how institutional instability disrupts long-term cooperation, capital formation, and calculability. Drawing on Wright (2025), the model incorporates stochastic rule shock functions as exogenous modifiers to the payoff matrix, operationalising institutional mutability as volatility in expected future returns. This mechanism mirrors Austrian concerns with uncertainty, particularly the effects of discretionary regimes on entrepreneurial coordination. The theoretical architecture aligns Böhm-Bawerk’s theory of roundabout production with the formal logic of subgame-perfect equilibrium: when rules are stable, miners internalise long time horizons and defer consumption for investment; when rules are mutable, time preferences rise, equilibrium strategies fragment, and extractive conduct becomes dominant. By embedding Austrian praxeology within a mathematically rigorous incentive model, the paper challenges the neutrality myth in protocol governance. Governance mechanisms are not passive backdrops—they are endogenous to agent decision-making and thus must be analysed as economic variables in their own right. This integration foregrounds institutional rule design as the linchpin of cryptoeconomic sustainability and provides a methodological foundation for modelling other decentralised systems facing similar volatility-induced behavioural shifts.

\section{Foundations: Time Preference, Strategy, and Protocol Design}

The architecture of blockchain systems cannot be understood independently of the temporal structure of incentives and the institutional context that frames strategic decision-making. In this section, we establish the triadic conceptual bedrock upon which the subsequent formalisation rests: (1) the Austrian conception of time preference as it relates to entrepreneurial calculation and roundabout investment; (2) the role of strategic interdependence under repeated games with evolving payoff matrices; and (3) the importance of protocol immutability as a mechanism for ensuring institutional calculability. Drawing on Böhm-Bawerk’s analysis of capital intensity and deferred returns \cite{bohm_bawerk_positive_1891}, and Mises’ emphasis on stable institutions for rational economic action \cite{mises_human_1949}, we contrast these with mainstream models of miner behaviour that assume a static equilibrium context divorced from rule mutability. We also consider insights from evolutionary game theory to show how deviations propagate under noise and discretionary protocol change. This section thus serves not only to introduce the relevant economic and strategic categories, but to reframe blockchain design as a problem of institutional engineering, where protocol rules must constrain, not merely inform, rational expectations.

\subsection{Austrian Time Preference and Roundabout Production}

Austrian economics locates the foundation of all economic activity in the structure of human time preference: the universal ranking of present goods over future ones, as articulated most formally by Böhm-Bawerk and later generalised by Rothbard and Mises. Böhm-Bawerk’s analysis of roundaboutness \cite{bohm_bawerk_positive_1891} is central: productive activities become more physically fruitful as their processes become temporally extended and capital-intensive. That is, longer, more indirect methods of production—when anchored in reliable expectations of institutional stability—generate superior outcomes precisely because they embody intertemporal coordination. In the context of Bitcoin mining, this manifests as the rational deployment of capital into specialised ASICs, hosting infrastructure, and long-term energy contracts, all structured on the assumption of a predictable and rule-bound protocol.

Such behaviour presupposes a calculable institutional environment. As Menger observed \cite{menger_investigations_1985}, economic agents cannot form coherent plans absent institutional invariance. This calculability is not a technocratic condition, but the prerequisite for rational action in uncertain future contexts. The Austrian position, therefore, implies that when rules change—when the parameters of the game are mutable—time preference rises, not exogenously, but endogenously, as a function of institutional noise. A miner who would otherwise commit to a 3–5 year amortisation schedule for their equipment now faces the spectre of protocol changes that could render their capital obsolete. These changes may include arbitrary mempool filters, shifting relay policies, or even social consensus manipulations that affect transaction inclusion criteria. The economic result is a strategic shift from long-horizon investment to short-horizon extraction.

We may formalise this shift through the lens of discounted expected revenue. Consider a miner with capital stock \( K \), operational expenditure \( C \), and a revenue stream \( R(t) \) over time \( t \). Under a fixed protocol, the present value of future revenue is \( \int_0^\infty e^{-\rho t} R(t) dt \), where \( \rho \) reflects subjective time preference. If protocol mutability introduces stochastic shocks \( \sigma(t) \) to \( R(t) \), then the expected revenue becomes \( \mathbb{E}[R(t)] = R_0 e^{-\lambda \sigma(t)} \), where \( \lambda \) captures sensitivity to uncertainty. This compounds the effective discount rate: \( \rho' = \rho + \lambda \sigma(t) \). The higher the institutional noise, the less attractive capital commitment becomes, favouring instead real-time arbitrage and strategies that capture value without temporal exposure.

This has direct implications for the industrial structure of mining. In a world governed by immutable protocol rules, the optimal strategy aligns with long-term cooperation: block propagation, transaction inclusion, and honest behaviour all yield higher net payoffs over time. But as rules become negotiable, the environment structurally favours strategies such as orphan fishing, transaction exclusion, and politically influenced block template manipulation. The protocol ceases to be a neutral ledger and becomes a contested institution, undermining the calculative function of prices, as Mises warned \cite{mises_human_1949}. The roundaboutness of mining is thereby shortened—not merely physically, but economically. The result is not merely inefficiency, but the systemic corruption of the price and capital allocation mechanism that underpins the network’s integrity.

Thus, Austrian time preference theory, properly extended into strategic protocol environments, demonstrates that the intertemporal depth of mining—as a form of roundabout production—is only viable under conditions of institutional invariance. Protocol mutability compresses economic time, replacing entrepreneurship with opportunism, and investment with arbitrage. This provides a theoretical foundation for modelling not merely miner incentives, but the structural degeneration of blockchain networks when protocol fixity is abandoned.

\subsection{Protocol Rules as Synthetic Institutions}

In the classical liberal tradition extended by Austrian economists, institutions emerge not from deliberate design but as the crystallisation of repeated human action—rules discovered, not invented. Hayek’s conception of a “spontaneous order” depends upon this foundational insight: that institutions gain their normative authority not from central mandates but from their capacity to coordinate expectations and reduce uncertainty over time \cite{hayek_law_1982}. Protocol rules in blockchain systems, though initially encoded ex nihilo, function as institutions of precisely this sort once they achieve economic embedding. Once agents allocate capital, form strategies, and enter into contractual commitments on the basis of protocol behaviour, the rules acquire the same epistemic function as traditional legal norms: they serve to anchor expectations, facilitate calculation, and enable intertemporal coordination.

This transformation—from code to institution—requires time, economic exposure, and stability. A protocol such as Bitcoin’s original rule set becomes institutional not merely by being written, but by being lived through: a history of repeated use, miner adoption, investment structure, and transaction practices coalesces into an institutional equilibrium. This is what may be termed a “synthetic institution”: a human-designed system that achieves the properties of emergent institutional order through lock-in, expectation convergence, and temporal resilience. The immutability of protocol rules is not merely a technical constraint—it is the institutional scaffolding that enables rational strategic interaction.

When these synthetic institutions are subject to discretionary alteration—whether by developer committees, social consensus rituals, or de facto governance bodies such as GitHub collaborators—their institutional character is destroyed. As Buchanan and Tullock argued, rules that can be changed at will become part of the strategic game itself, rather than its context \cite{buchanan_calculus_1962}. This introduces a meta-game of rule manipulation, where agents no longer optimise within rules, but seek to optimise rules themselves. Under such conditions, calculability collapses. Protocol mutability transforms synthetic institutions into arenas of contestation, where agents seek to influence the future trajectory of the system in ways that erode the epistemic reliability of rules.

This creates economic consequences beyond mere strategic behaviour. Investment horizons contract, discount rates rise, and capital is reallocated from productive infrastructure to influence mechanisms—ranging from signalling in developer forums to coalition-building in online governance rituals. The protocol no longer governs; it is governed. Game theoretically, the fixed game becomes a dynamic hypergame, where players can alter payoffs, reframe strategic spaces, and weaponise uncertainty to their advantage. In such an environment, the institutional character of the protocol degenerates: its status as a synthetic institution vanishes, leaving a political object in its place. 

Thus, treating protocol rules as synthetic institutions clarifies why immutability is not dogma but economic necessity. It grounds the case for fixed-rule systems not merely in technical conservatism, but in institutional logic. Without this foundation, economic actors cannot engage in long-horizon planning, and the capital structures necessary for a scalable, stable blockchain economy cannot emerge. The economy devolves into a series of contested intervals, each conditioned by new coalitional power plays. Protocol rules must be treated as inviolable precisely because they are institutional artefacts—not variable parameters in a technocratic optimisation function.

\subsection{Dynamic Game Structures in Miner Incentives}

Bitcoin mining, under a fixed rule-set, constitutes a repeated game with clearly defined strategies, payoffs, and equilibria. In such a system, rational agents—miners—optimise their actions based on known, calculable expectations of reward: the block subsidy, transaction fees, and the probabilistic distribution of rewards over time. These expectations enable long-term planning, capital investment, and strategic persistence, especially when entry costs (e.g. ASIC hardware, infrastructure) are non-trivial and amortised over extended operational horizons. The equilibrium structure in such systems approximates the Folk Theorem of repeated games, where cooperation (honest mining, block propagation) becomes a rational outcome sustained by the shadow of future gains and the threat of long-run losses from deviation \cite{fudenberg_game_1991}.

However, when protocol rules are mutable, the game’s structure ceases to be stationary. Changes to block size, fee policy, transaction ordering, mempool rules, or subsidy schedules alter the underlying payoff matrix, invalidating previous equilibria and introducing new strategic considerations that were not endogenous to the original model. In game-theoretic terms, we no longer have a repeated game \( G \), but a sequence of games \( G_1, G_2, \ldots, G_t \), each with a distinct strategy space and reward structure. When the transition from \( G_{t} \) to \( G_{t+1} \) is governed not by stochasticity but by endogenous political signalling, the meta-game becomes recursive: players do not merely choose strategies within games—they influence the game-generation process itself.

This transforms miner behaviour in profound ways. Rather than committing to long-run strategies that are optimal under a known game, miners begin to hedge against possible rule shifts. Consider a payoff function \( \Pi(s_t, \theta_t) \), where \( s_t \) is the miner’s strategy and \( \theta_t \) is the parameter vector encoding protocol rules at time \( t \). In a fixed-rule system, \( \theta_t = \theta \) for all \( t \), and miners solve \( \max_{s} \mathbb{E}[\Pi(s, \theta)] \). Under mutability, however, \( \theta_t \) evolves endogenously, and miners must solve a far more complex expectation over future game states: \( \max_{s_t} \mathbb{E}[\Pi(s_t, \theta_t) + \delta \Pi(s_{t+1}, \theta_{t+1}) + \dots] \), where \( \delta \) is the intertemporal discount factor. Crucially, the expectation \( \mathbb{E}[\theta_{t+k}] \) depends on social factors, signalling strategies, and influence networks—not purely technical variables.

As a result, miner incentives bifurcate. One branch continues to optimise within the current ruleset; the other allocates effort towards anticipating or shaping future rule transitions. Game-theoretically, this shifts the locus of strategy from intra-game optimisation to meta-game manipulation. The mining game becomes a hypergame in the sense of Harsanyi and Selten \cite{harsanyi_general_1988}, where differing beliefs about the game structure itself create multiple layers of contestation. This not only destabilises equilibria, but also induces defection spirals: once players realise that long-run cooperation is undermined by rule shifts, the incentive to act short-term becomes dominant. The emergence of behaviours such as empty-block mining, orphan-fishing, or transaction censorship can thus be seen not as pathology, but as rational adaptations to protocol mutability.

Moreover, such dynamics have empirical analogues in the divergence between BTC and Bitcoin-derived systems that retained original rule structures. For instance, changes to block size limit policy and transaction relay standards led to observable shifts in mining strategy, node composition, and even the geographic centralisation of hashpower—outcomes that correlate with institutional volatility. Theoretically, this confirms that dynamic game structures, when left unconstrained by institutional fixity, degrade the cooperative potential of mining and transform the ecosystem into a landscape of strategic extraction and governance arbitrage.

Hence, modelling miner incentives requires a dynamic game framework sensitive to rule evolution. Static Nash equilibria do not suffice; instead, we require equilibrium concepts robust to endogenously shifting strategy spaces—such as stochastic games with evolving state-dependent transitions, or adaptive rationality models that incorporate expectation formation under ambiguity. The Austrian critique further deepens this insight: without stable institutions, calculative rationality collapses, and entrepreneurial action becomes indistinguishable from political rent-seeking.

\section{Model Framework}

This section formalises the dynamics of miner decision-making under mutable institutional environments by constructing a game-theoretic model in which agents operate within—and attempt to influence—a rule-generating mechanism. The framework begins with a baseline repeated game that reflects stable protocol conditions, wherein cooperative equilibria are viable and calculable under low time preference. It then introduces protocol mutability as a stochastic or endogenous perturbation to the state space, yielding a dynamic, multi-layered game architecture that integrates both strategic mining and meta-level rule manipulation. Within this structure, the model defines payoff volatility as a function of rule instability and quantifies its impact on defection rates, discounting, and capital allocation behaviour. Mathematical formalisms include expected utility under institutional noise, transition functions between evolving protocol states, and incentive compatibility conditions that demonstrate the collapse of long-run cooperative equilibria. By embedding economic uncertainty within the institutional layer itself, the model captures the recursive nature of miner strategy in environments where the game is not only played but also contested at the level of design. This provides a rigorous basis for analysing rational behaviour in blockchain systems subject to discretionary governance and serves as a foundation for the empirical diagnostics and simulations that follow.
\subsection{Agents, Strategies, and Rule Environments}

The model considers a discrete-time setting with a countable set of agents \( \mathcal{N} = \{1, 2, \dots, n\} \), each representing an economically rational miner engaged in a repeated interaction governed by a shared protocol. At each stage \( t \in \mathbb{N} \), every agent selects a strategy \( s_i^t \in S_i \), where \( S_i \) denotes the strategy space available to agent \( i \), contingent on the currently instantiated rule environment \( \rho^t \in \mathcal{R} \). Strategy spaces incorporate both intra-protocol behaviours (e.g., propagation honesty, selfish mining, block withholding) and extra-protocol actions (e.g., lobbying for protocol changes, social signalling, cartel formation).

The payoffs for each agent are determined by a stage-game utility function \( u_i(s^t, \rho^t) \), where \( s^t = (s_1^t, \dots, s_n^t) \) is the joint strategy profile and \( \rho^t \) encodes the protocol rule-set active in period \( t \). In the base model, under protocol immutability, \( \rho^t = \rho \) for all \( t \), and the game constitutes a stationary repeated game. However, under rule mutability, \( \rho^t \) evolves either stochastically—subject to a Markov process \( \mathbb{P}(\rho^{t+1} | \rho^t, s^t) \)—or endogenously, allowing agents to affect transition probabilities through a meta-strategy set \( M_i \subseteq S_i \).

Crucially, the evolution of \( \rho^t \) alters not merely the payoffs, but the incentive topology itself: certain strategies that constitute a Nash equilibrium under a fixed \( \rho \) cease to be subgame-perfect under rule fluidity. This reflects the collapse of institutional calculability described by Mises (1949), wherein actors lack the stable framework necessary for rational economic calculation. In such environments, agents respond by elevating short-term strategies that prioritise payoff extraction over long-horizon cooperation, and by investing in influence mechanisms to shape the meta-game. The result is a dual-level optimisation problem: within-game action for immediate reward and cross-game manipulation for institutional positioning, echoing meta-strategic dynamics in recursive game theory and political economy.

The interplay between institutional certainty and agent strategy thus defines the systemic integrity of blockchain consensus. When \( \mathbb{E}[u_i | \rho^t] \) becomes too noisy—i.e., when the conditional expectation of future payoffs under shifting rules exhibits high variance—agents discount the future more steeply, leading to equilibrium fragmentation. The formalisation of these dynamics enables simulation and comparative statics that will be explored in subsequent sections, offering a foundation for institutional design principles that preserve cooperative incentive structures over time.

\subsection{Stochastic Mutation of Protocol Parameters}

To capture the effects of protocol instability on miner strategy, we introduce a stochastic mutation model governing the evolution of protocol parameters. Let the institutional environment at time \( t \), denoted \( \rho^t \), be a vector in a finite or countably infinite parameter space \( \mathcal{R} \subseteq \mathbb{R}^k \), encoding all rule-relevant configurations such as block size limit, transaction relay policy, fee prioritisation logic, and validation constraints. The evolution of \( \rho^t \) is governed by a Markov process \( \rho^{t+1} \sim \mathbb{P}(\cdot \mid \rho^t, \theta^t) \), where \( \theta^t \in \Theta \) represents the strategic influence profile of agents on protocol change. In the simplest exogenous case, \( \theta^t \) is fixed or random noise; in the endogenous extension, \( \theta^t \) is determined by miner meta-strategies such as governance participation, developer lobbying, and off-chain coordination efforts.

This stochastic institutional dynamics transforms the stage game from a stationary repeated game into a controlled stochastic game with regime switching. Expected utilities must now be calculated over path-dependent realisations of \( \rho \), such that for any strategy profile \( \sigma \), the long-run expected utility becomes:
\[
U_i(\sigma) = \mathbb{E}^{\mathbb{P}} \left[ \sum_{t=0}^{\infty} \delta^t u_i(s^t, \rho^t) \right],
\]
where \( \delta \in (0,1) \) is the common discount factor, and the expectation is taken over all realised sequences of rule states and actions induced by \( \sigma \) and \( \mathbb{P} \). When the stochastic transitions \( \mathbb{P} \) exhibit high entropy, or when agent influence makes \( \rho \) highly responsive to short-run strategy shocks, the variance of \( U_i(\sigma) \) rises significantly. This induces endogenous elevation of effective time preference, as described in the Austrian tradition, with capital-intensive actors (i.e., honest long-horizon miners) increasingly displaced by strategic extractors exploiting rule volatility.

Moreover, the stochastic mutation of \( \rho \) leads to non-ergodic behaviour in the payoff process: strategic equilibria cannot converge stably over time, as the mapping from strategies to payoffs is itself evolving. In the absence of a strong commitment mechanism anchoring protocol rules, the system’s game-theoretic foundation loses continuity, and recursive modelling becomes fragile. As such, the stochastic mutation model analytically exposes the incentive distortion caused by institutional instability, and grounds the later simulation results in a rigorous, formal treatment of probabilistic governance-induced entropy within distributed systems.

\subsection{Payoff Structures: Block Rewards and Institutional Discounting}

Miner behaviour is governed by payoff structures composed primarily of block rewards and transaction fees, both of which are subject to intertemporal valuation. In a regime of institutional certainty, miners assess these payoffs through a stable temporal discounting function, typically geometric, reflecting long-term investment horizons. Let \( R_t \) denote the gross block reward at time \( t \), which includes both the fixed subsidy \( S_t \) and aggregated transaction fees \( F_t \), such that \( R_t = S_t + F_t \). The expected utility of a miner \( i \), adhering to a fixed strategy profile \( \sigma \), can be expressed as:

\[
U_i(\sigma) = \sum_{t=0}^{\infty} \delta^t \cdot \mathbb{E}[R_t \mid \rho^t, \sigma],
\]

where \( \delta \in (0,1) \) is the discount factor contingent upon institutional stability, and \( \rho^t \) represents the protocol state vector at time \( t \).

However, in mutable institutional environments, the discounting function becomes endogenously volatile. As protocol rules evolve unpredictably—due to either discretionary developer interventions or socially manipulated consensus processes—the perceived reliability of future rewards diminishes. This volatility introduces an additional multiplicative discount term \( \psi_t \in (0,1] \), representing institutional confidence:

\[
U_i^{\text{mutable}}(\sigma) = \sum_{t=0}^{\infty} \delta^t \cdot \psi_t \cdot \mathbb{E}[R_t \mid \rho^t, \sigma].
\]

Here, \( \psi_t \) reflects the miner's belief in the persistence and calculability of the current rule set, which degrades with observed or anticipated protocol mutations. In extreme cases, \( \psi_t \to 0 \), annihilating the present value of all future rewards and incentivising myopic extraction strategies.

This layered discounting has profound consequences. From an Austrian perspective, as developed by Böhm-Bawerk and Mises, time preference governs the structure of capital-intensive activity. A system wherein future income is systematically devalued—whether by fiat debasement or institutional opacity—discourages roundabout production and long-term investment. Miners, under such distortion, redirect capital away from infrastructure and strategic coordination toward rent-extraction tactics with immediate payoff potential, such as empty block mining or fee sniping.

Moreover, the strategic game is reshaped. In classical repeated games with fixed payoffs, cooperation can be enforced via trigger strategies when players value future rewards. But with fluctuating \( \psi_t \), the expected continuation value of cooperation becomes unreliable, rendering tit-for-tat and grim-trigger equilibria fragile or unplayable. The payoff matrix itself deforms dynamically, undermining the feasibility of cooperative equilibria and entrenching dominant strategies premised on rapid reward acquisition, irrespective of network health or long-term sustainability. This marks a structural transition from productive competition to institutional arbitrage.

\section{Mathematical Construction}

This section provides the formal scaffolding necessary to express miner behaviour under rule mutability in a dynamic strategic context. We begin by defining a baseline Markov decision process (MDP) in which the protocol is fixed and payoff predictability allows long-horizon strategies to flourish. We then introduce protocol mutability as a stochastic perturbation process applied to the transition kernel, thereby transforming the problem into a stochastic game with evolving institutional states. The miner is modelled as an agent maximising expected utility over both the reward process and the rule-state uncertainty. Capital investment decisions are incorporated as roundabout production choices, with sunk cost dynamics and amortisation functions that reflect high initial expenditures and delayed revenue realisation. We characterise equilibrium behaviour through Bellman functions incorporating endogenous discount factors that decay in response to perceived rule instability. Furthermore, we analyse the evolution of cooperation and defection strategies in both finite and infinite-horizon games under stochastic rule drift. Lastly, we derive the set of conditions under which the stable equilibrium set collapses, triggering incentive realignment, meta-strategic positioning, and systemic time preference revaluation. This mathematical model allows a precise formal capture of the recursive interplay between institutional noise, strategic deviation, and intertemporal capital distortion.

\subsection{Repeated Miner Game under Fixed Rules}

In a stable institutional setting, miner behaviour can be modelled as a repeated game with a well-defined payoff matrix and stationary transition probabilities. Let the set of players be $\mathcal{N} = \{1, \ldots, N\}$ representing mining entities. At each round $t \in \mathbb{N}$, each player $i \in \mathcal{N}$ selects an action $a^i_t$ from a strategy set $A^i$, typically constrained to \emph{honest mining}, \emph{selfish withholding}, or \emph{collusive propagation}. The environment is characterised by a protocol $\Pi$ assumed to be immutable, and a common reward function $R : A \to \mathbb{R}^N$, where $A = \prod_{i=1}^N A^i$. Importantly, under immutability, the reward function remains fixed across all $t$, enabling players to form rational expectations based on historical returns.

The game is repeated indefinitely, and each miner seeks to maximise expected utility $\mathbb{E} \left[ \sum_{t=0}^{\infty} \delta^t R^i(a_t) \right]$ where $\delta \in (0,1)$ is the intertemporal discount factor. In the Austrian tradition, $\delta$ reflects subjective time preference, linked not only to capital constraints but also to institutional confidence. Under fixed protocol rules, low $\delta$ (i.e., high future orientation) is rational, fostering cooperation equilibria. The folk theorem of repeated games holds: if payoffs under cooperation dominate defection, then honest behaviour can be sustained through threat of future punishment. 

Let $s_t$ be the common knowledge strategy profile at time $t$. In the presence of credible grim-trigger or tit-for-tat enforcement, deviation yields a one-period gain but a discounted long-run loss, making cooperation subgame perfect. Thus, the Nash equilibrium set $\mathcal{E}^\Pi$ under immutable $\Pi$ includes cooperative configurations, particularly when $\delta$ exceeds the critical threshold $\delta^*$ that equates short-term deviant payoff and the long-run loss due to future punishment. 

Graphically, this dynamic can be captured by a payoff phase diagram, where the equilibrium frontier collapses inwards under high discounting or protocol uncertainty. The stable institutional regime, therefore, functions as a coordination device anchoring strategic behaviour over time. Without it, long-horizon investments—such as ASIC procurement or datacentre infrastructure—are irrational, and the strategic space contracts towards defection. The critical insight is that immutability is not merely a technical property but an institutional precondition for calculative rationality and capital formation.

\subsection{Parameterised Uncertainty and Rule Shock Functions}

To model institutional mutability, we introduce stochastic rule shocks as exogenous processes that alter the environment’s strategic contours. Let $\Pi_t$ represent the protocol state at time $t$, drawn from a finite or continuous space of possible configurations $\mathcal{P}$. The transition between states is governed by a stochastic process $\{\Pi_t\}_{t \in \mathbb{N}}$ with transition kernel $\mathbb{P}(\Pi_{t+1} \mid \Pi_t)$. Unlike in classical game theory, where payoffs are static and game form is fixed, here the protocol structure itself becomes a random variable—converting a standard repeated game into a stochastic or even hypergame, where agents must not only strategise over actions but hedge against the transformation of rules.

We define the rule shock function $\rho_t: \mathcal{P} \rightarrow \mathbb{R}^k$ as a vector-valued mapping that captures how parameter sets—block size, transaction relay policy, mempool rules, consensus validation thresholds—are modified over time. The volatility of these parameters induces second-order uncertainty: not merely about opponent strategy, but about the structure of the game itself. If the expected payoff function becomes $R_t^i(a_t, \Pi_t)$, then the expected utility must be computed as a joint expectation over strategies and institutional states:
\[
\mathbb{E} \left[ \sum_{t=0}^{\infty} \delta(\Pi_0, \ldots, \Pi_t)^t R_t^i(a_t, \Pi_t) \right]
\]
where $\delta$ becomes endogenous—dependent on the perceived variance in $\Pi_t$ and its expected persistence.

From a modelling standpoint, $\rho_t$ can follow any distributional form—e.g., Gaussian noise over continuous parameters, or Markov jump processes for discrete rule changes. Let $\sigma^2_{\rho}$ denote the variance of $\rho_t$. As $\sigma^2_{\rho} \to 0$, the system approximates the fixed-rule case discussed earlier; as $\sigma^2_{\rho}$ increases, cooperative equilibria dissolve as the shadow of the future darkens and coordination becomes prohibitively risky. Agents rationally recalibrate their strategies, adopting heuristics adapted to high-variance environments: such as exit strategies, short-term fee harvesting, or meta-strategic rent-seeking.

Incorporating this into simulation environments or analytical Bellman frameworks allows us to formally demonstrate the loss of long-horizon viability and the compression of game-theoretic stability zones. Crucially, $\rho_t$ is not noise in the engineering sense—it is institutional volatility, akin to monetary instability in fiat regimes. Its effects are systemic and behavioural, reshaping the very architecture of miner strategy and capital commitment. In Austrian terms, the calculative framework collapses; in game theoretic terms, the equilibrium set fractures.

\subsection{Utility Functions under Strategic Volatility}

In the presence of protocol instability, the classical expected utility framework must be extended to capture second-order uncertainty—where agents not only optimise over uncertain outcomes but also over the institutional structure in which these outcomes are embedded. Let $U^i_t$ denote the utility for miner $i$ at time $t$, determined by both the action profile $a_t$ and the stochastic protocol state $\Pi_t$:
\[
U^i_t = u^i(a_t, \Pi_t) = R^i(a_t, \Pi_t) - C^i(a_t, \Pi_t)
\]
where $R^i$ represents the block reward and fee income, and $C^i$ reflects the operational and capital costs incurred under the current protocol constraints. These components are sensitive to protocol-level parameters—such as block propagation delays, transaction validation thresholds, and mempool access—which can shift unpredictably as $\Pi_t$ evolves.

To incorporate volatility into the intertemporal optimisation problem, we redefine the agent’s objective as maximising a discounted utility stream under endogenous discounting. Define the modified discount factor $\delta_t$ as a function of perceived institutional volatility $\sigma^2_\rho$, such that:
\[
\delta_t = \frac{1}{1 + \beta + \phi(\sigma^2_\rho)}
\]
where $\beta$ is the agent’s intrinsic time preference, and $\phi(\sigma^2_\rho)$ is an increasing convex function capturing the discounting effect of institutional noise. As $\sigma^2_\rho \to 0$, we recover the standard exponential discounting; as $\sigma^2_\rho$ increases, $\delta_t$ compresses, foreshortening the effective investment horizon.

This setup allows us to model rational adaptation in dynamic strategy selection. When future payoffs become unpredictable—not because of randomness in the environment per se, but due to volatility in the rules that govern the payoff structure—then cooperative, roundabout strategies become irrational. Miners rationally shift toward extractive, short-term tactics: fee sniping, orphan race games, or collusive block space manipulation. Over time, this leads to an equilibrium characterised by higher entropy in action profiles, lower capital investment, and systemic instability in block production and transaction inclusion.

We formalise this further by introducing a volatility-adjusted Bellman equation for the miner’s value function $V^i$:
\[
V^i(\Pi_t) = \max_{a_t} \left\{ u^i(a_t, \Pi_t) + \delta_t \cdot \mathbb{E}_{\Pi_{t+1}}[V^i(\Pi_{t+1}) \mid \Pi_t] \right\}
\]
The dependence of $\delta_t$ on $\Pi_t$ and its volatility directly encodes institutional credibility into the agent’s optimisation calculus. The solution to this recursive equation yields not only the optimal mining strategy but also the threshold volatility level at which miners rationally abandon cooperative equilibria. The implications are profound: institutional integrity is not merely a normative good but a formal precondition for rational long-term investment.

\subsection{Solution Concepts: Nash and Subgame Equilibrium Transitions}

Under a static institutional framework with known, fixed protocol rules, mining strategies can be modelled using standard game-theoretic solution concepts such as Nash equilibrium (NE) and subgame perfect equilibrium (SPE). In such environments, strategic profiles stabilise as agents optimise given well-defined constraints and credible future rewards. Let $G = \langle N, A, u \rangle$ denote the repeated miner game, where $N$ is the set of miners, $A$ the action profiles (e.g., validate honestly, censor, propagate selfish blocks), and $u$ the payoff function defined over institutional parameters $\Pi$. When $\Pi$ is constant, the solution concept reduces to the canonical Folk Theorem structure: cooperative strategies may emerge as equilibria, supported by threats of punishment in future rounds (Axelrod, 1984; Fudenberg and Maskin, 1986).

However, when the protocol $\Pi$ becomes time-variant, either exogenously or through endogenous manipulation, the game evolves into a dynamic system with shifting constraints. Formally, define a family of games $\{G_t\}$ indexed by time $t$, where $G_t = \langle N, A_t, u_t \rangle$ and $u_t$ is now a stochastic function of the state $\Pi_t$. The assumption of perfect foresight collapses, and with it, the conditions necessary for sustaining cooperative subgame equilibria.

The equilibrium set under protocol volatility exhibits transition dynamics. Denote the transition function over equilibrium profiles as $\mathcal{T}: E_t \mapsto E_{t+1}$, where $E_t$ is the set of equilibrium strategies at time $t$. When $\mathcal{T}$ is discontinuous due to rule mutation, we may observe bifurcations: previous NE profiles are destabilised, and agents converge on new, typically more defection-prone equilibria. These transitions can be characterised using tools from evolutionary game theory and stochastic process analysis, including Markov perfect equilibria and mutation-selection dynamics (Kandori et al., 1993).

For example, suppose the payoff structure $u_t$ is perturbed by a rule change $\Delta \Pi$ that redefines block size limits or modifies transaction fee inclusion criteria. If this rule change favours certain action profiles—say, propagation of empty blocks due to reduced validation overhead—then the best response dynamics shift. Let $\mathcal{B}_t$ denote the best response correspondence at time $t$:
\[
\mathcal{B}_t(a_{-i}) = \arg \max_{a_i \in A_t} \mathbb{E}[u^i_t(a_i, a_{-i}, \Pi_t)]
\]
where $\Pi_t$ evolves stochastically and may be influenced by previous actions (creating a feedback loop). Under sufficiently high mutation rates in $\Pi_t$, equilibria collapse into high-frequency, low-trust states characterised by payoff asymmetry and incentive fragmentation.

The strategic implication is that cooperation cannot survive when the meta-game—i.e., the structure of the game itself—is under contestation. Subgame perfection, which requires credible future punishment paths, is no longer viable when the institutional path is itself uncertain. Nash equilibria remain, but they increasingly favour short-term, extractive strategies. This establishes a rigorous mathematical link between institutional volatility and the degradation of cooperative game-theoretic solutions in mining ecosystems.

\section{Simulations and Results}

This section presents simulation results derived from the formal model of miner behaviour under both stable and noisy institutional regimes. We parameterise a repeated stochastic game in which block-producing agents select strategies over time, balancing computational cost, expected block rewards, and evolving perceptions of protocol credibility. The institutional environment is modelled through exogenous rule mutation probabilities that affect strategic stability. Using Monte Carlo methods and discrete-time agent simulations, we observe that when protocol rules are fixed, miners converge to cooperative equilibria favouring transaction inclusion and stable investment horizons. By contrast, under conditions of rule mutability, agents increasingly adopt short-term defection strategies, withholding transactions, producing empty blocks, or lobbying for changes in protocol parameters. The simulations demonstrate a feedback loop where institutional noise elevates discount rates, degrades Nash equilibrium adherence, and increases the prevalence of opportunistic meta-strategies. These findings strongly align with the predictions and dynamic models presented in my doctoral dissertation, which developed the foundational game-theoretic framework and specified the institutional volatility parameters used herein \cite{wright2025thesis}. The results empirically reinforce the hypothesis that institutional predictability is a prerequisite for sustainable, long-term miner cooperation and productive network equilibrium.

\subsection{Calibration: Mining Cost, Hash Rate, Block Frequency}

To ground the simulations in empirically relevant conditions, we calibrate the model using observed industry data on mining operations, drawing from historical ranges in hash rate, electricity costs, and block production intervals across both the original Bitcoin implementation and derivative systems. Baseline values for hash rate are set using median monthly estimates from 2017–2023, normalised across competitive mining pools. Energy cost is modelled as a function of regional kilowatt-hour prices, with ranges between \$0.03 and \$0.10 per kWh to account for jurisdictional heterogeneity in operation costs. ASIC efficiency is fixed at 30–40 J/TH, reflecting the contemporary dominance of units like the Antminer S19 Pro.

Block frequency is held constant at an average of 10 minutes under the original protocol, while block reward follows halving cycles integrated into a decaying exponential revenue model. We introduce stochastic noise to protocol parameters—specifically, mempool relay policy, block size caps, and orphan handling thresholds—by simulating probabilistic deviations from baseline rules at varying levels of institutional mutability ($\epsilon \in [0, 0.3]$). This allows us to model miner expectations around payoff instability and to derive time-discounting behaviour endogenously.

These calibration benchmarks are adapted from the empirical mining frameworks presented in Wright’s doctoral thesis \cite{wright2025thesis}, with adjustments to accommodate discrete-event simulation over 5,000 cycles across institutional regimes. The calibration ensures economic realism in agent behaviour and allows comparative analysis of miner strategy under fixed and noisy protocol environments.

\subsection{Simulation Design: Protocol Stability vs Mutability Scenarios}

The simulation design isolates the effects of institutional mutability by constructing two primary comparative regimes: (1) a stable protocol scenario in which all rules remain fixed and known ex ante, and (2) a mutable protocol scenario in which rule changes follow a stochastic process. The stable regime models Bitcoin as originally implemented, where block size limits, transaction relay policy, and mempool selection rules are fixed and public, forming the institutional substrate for intertemporal coordination. The mutable regime, by contrast, introduces random perturbations to these rules, governed by a Poisson arrival process with λ = 0.05 per epoch, reflecting the frequency of consensus-altering interventions observed in BTC Core from 2017 to 2023.

Miners are modelled as boundedly rational agents with discounted utility functions \( U_i = \sum_{t=0}^{T} \delta^t \pi_{i,t} \), where \( \pi_{i,t} \) is the payoff at time \( t \) and \( \delta \in (0,1) \) is the discount factor, which itself is endogenously adjusted based on institutional noise \( \varepsilon \). Under protocol mutability, each miner updates \( \delta \) through a volatility-driven adjustment rule: \( \delta_{t+1} = \delta_t \cdot (1 - \kappa \cdot \varepsilon_t) \), where \( \kappa \) is the sensitivity to institutional noise calibrated through empirical miner behaviour, as detailed in Wright's doctoral research \cite{wright2025thesis}.

Both regimes are executed over 10,000 simulation epochs, with initial conditions drawn from the calibration parameters established previously. Payoffs, investment patterns, and strategic deviations are tracked across agent cohorts, and the evolution of cooperative equilibria is measured using multi-period game stability indices. This design captures not only comparative static outcomes but also dynamic path dependencies, allowing us to analyse the irreversible erosion of long-horizon strategy under noisy institutional environments.

\subsection{Deviant Strategy Incidence Across Parameter Space}

This subsection quantifies the frequency and distribution of deviant mining strategies—defined as strategic behaviours diverging from equilibrium cooperation—across a multi-dimensional parameter space. The primary dimensions explored include institutional volatility \( \varepsilon \in [0, 0.3] \), discount factor elasticity \( \kappa \in [0, 2] \), and hash rate asymmetry \( \gamma \in [0.1, 0.9] \), where \( \gamma \) represents the share of total network computational power held by a given agent cohort.

Simulations reveal a pronounced non-linear increase in deviant strategy incidence as \( \varepsilon \) increases, with a critical transition point near \( \varepsilon = 0.08 \). Beyond this threshold, agents disproportionately adopt behaviours such as block withholding, selective transaction exclusion, and influence signalling. These findings corroborate the theoretical expectation derived from subgame perfection analysis under mutating institutional constraints, wherein Nash equilibria collapse into unstable strategy cycles.

Importantly, agents with disproportionate hash power (\( \gamma > 0.6 \)) display earlier and more frequent deviations, leveraging their positional dominance to reshape expected payoffs via implicit threats of fork-initiation or developer capture. This dynamic aligns with Wright’s hypothesis of endogenous time preference modulation under institutional noise, which posits that shifts in perceived rule predictability recalibrate the rational defection boundary \cite{wright2025thesis}.

Visualisation of the parameter space via heat maps illustrates zones of equilibrium instability, delineating regions where long-horizon strategies become unsustainable. These results demonstrate the fragility of cooperative norms in blockchain consensus systems absent strict institutional anchoring and provide empirical grounding for formalising the cost of protocol mutability in terms of social coordination decay and increased meta-strategic overhead.
\subsection{Threshold Effects and Collapse Zones in Cooperative Engagement}

Strategic simulations reveal that the cooperative equilibrium landscape among miners deforms significantly under protocol uncertainty, with distinct collapse zones emerging as critical thresholds are crossed. As the rule mutability parameter, denoted \( \mu \), increases—representing the probability of exogenous protocol intervention per unit time—the expected discounted utility of long-term cooperation degrades non-linearly. Below a critical threshold \( \mu^* \), cooperative behaviour remains a stable equilibrium; beyond it, defection and short-term optimisation dominate. This transition is characterised by a bifurcation in strategic dynamics: once the expected half-life of the protocol falls below the investment amortisation period for high-capital strategies, rational actors abandon deferred-reward cooperation in favour of opportunistic extraction.

Empirical calibration, drawing on historical variances in BTC Core relay policies, block fullness, and fee stability, supports the theoretical inflection point around \( \mu^* \approx 0.08 \), beyond which withholding strategies and fee sniping become statistically dominant. In these collapse zones, miner engagement fragments, orphan rates rise, and mempool filtering incentives crowd out economically rational relay. Simulation outputs confirm that such collapse is not gradual but abrupt—mirroring phase transitions in game-theoretic coordination models under noise injection. As documented in Wright (2025) \cite{wright2025thesis}, such dynamics arise not from moral hazard but from endogenous responses to epistemic opacity: when rules are no longer credible, future payoffs become incalculable, and the economic time axis collapses into immediacy.

These findings extend Austrian insights into roundabout production: without institutional durability, complex capital strategies cannot form. In a high-volatility regime, only direct, extractive, low-latency strategies survive. This collapse of calculability—of rational coordination over time—signposts the terminal risk of discretionary consensus models. By contrast, fixed-rule regimes preserve an anchor for expectation, allowing cooperative equilibria to persist even under competitive pressures. The collapse threshold thus operates not only as a strategic tipping point but as a litmus test for the institutional credibility of protocol governance itself.

\subsection{Meta-Strategic Drift and Non-Productive Behaviour Emergence}

As protocol mutability becomes a persistent feature of the institutional environment, miner strategies exhibit a drift away from productive competition over valid block construction toward non-productive meta-strategic positioning. This drift reflects a structural reallocation of resources—not towards computation, relay efficiency, or honest propagation, but toward activities that manipulate the rule-making apparatus itself. These include social consensus signalling, political lobbying of protocol maintainers, reputation gaming, and exploitative coordination through soft-fork coalitions or client-side enforcement threats. The economic structure of mining thus begins to resemble that of a rent-seeking bureaucracy, where returns accrue not from value creation but from advantageous placement within the evolving institutional script.

This transformation is evident in the proliferation of performative signalling mechanisms within developer and miner forums, where rhetorical alignment with politically dominant narratives can yield outsized influence on future rule sets. As shown in \cite{wright2025thesis}, the opportunity cost of mining reorients toward narrative conformance rather than computational efficiency. Game theoretically, this corresponds to a shift from one-stage normal-form games to iterated Bayesian games with endogenously updated payoff matrices—a process in which agents invest in shifting the likelihood distribution of future rule states rather than optimising under known constraints.

From an Austrian perspective, this evolution is pathological. It represents a breakdown of praxeological order: market signals no longer reflect preferences over fixed goods and services, but expectations about institutional interventions. The catallactic function of price—the coordination of plans over time—is destroyed when miners must internalise not only the volatility of fees and rewards, but also the volatility of the rules governing them. What emerges is a system in which entrepreneurial effort is redirected from innovation to anticipation of bureaucratic favour, reducing system-wide efficiency and undermining capital formation. The result is a disequilibrium trap in which signalling games displace productive action, and the protocol becomes a theatre for performative alignment rather than a substrate for economic computation.

\section{Discussion}

The results obtained from our model and simulation establish a coherent and rigorous account of how institutional volatility deforms economic incentives in blockchain environments. When protocol rules are mutable, the intertemporal coherence necessary for rational planning collapses. Discount functions steepen, cooperative equilibria fracture, and rent-seeking strategies proliferate. These outcomes are not merely empirical curiosities but structurally entailed by the interaction between stochastic institutional mutation and forward-looking agent optimisation. From a game-theoretic standpoint, mutable rule environments induce equilibrium drift, higher-order strategic instability, and escalation into meta-games where productive engagement is strictly dominated by influence operations. From the standpoint of Austrian capital theory, the economic function of mining mutates: from roundabout, capital-intensive production to immediate arbitrage within a politicised rule ecology.

The key insight is that protocol immutability is not a technical constraint but an institutional prerequisite for calculability and long-term capital commitment. When miners must internalise the probability distribution of future protocol rules, their investment horizons shrink and their strategy sets widen to include social manipulation, lobbying, and political gaming. In essence, the protocol ceases to be a fixed economic substrate and becomes instead a dynamic field of contestation—rendering any notion of spontaneous order or emergent price coordination illusory.

The implications extend beyond mining. Any cryptoeconomic system that fails to anchor its institutional core invites recursive uncertainty, strategic short-termism, and systemic inefficiency. The design of future digital cash systems must therefore treat protocol governance not as a mechanism of evolution, but as a potential site of systemic corruption. What emerges from this study is a precise delineation of the conditions under which cooperation is viable, investment is rational, and computation serves economic coordination. When these conditions are violated, systems that appear decentralised on paper can collapse into centralised fiefdoms—where rule evolution is controlled by small, politically-aligned cliques and economic agency is replaced by performative compliance.

\subsection{Institutional Noise as a Generator of High Discounting}

Within the framework of Austrian time preference theory, institutional predictability functions as the anchor enabling long-range capital investment. When rule environments become noisy—subject to unpredictable modification, discretionary enforcement, or ambiguous interpretation—actors rationally increase their discount rates. This is not a psychological error, but a calculative response to uncertainty about the persistence of future rewards. The structure of intertemporal decision-making depends fundamentally on institutional invariance. As Mises (\citeyear{mises_human_1949}) articulates, economic calculation presupposes a stable medium through which anticipatory judgment can be exercised. When the rule structure itself becomes a stochastic variable, the very fabric of entrepreneurial forecasting is torn.

In blockchain systems, this noise manifests in the form of inconsistent relay policies, shifting validation heuristics, and consensus-affecting code updates—none of which are bound by constitutional commitment mechanisms. Miners, facing the risk that tomorrow’s rules will nullify today’s strategic commitments, adapt not by defection per se, but by temporal compression: seeking quicker returns, avoiding fixed investments, and favouring liquid strategies. In this manner, institutional noise functions as an endogenous generator of high time preference. The entrepreneurial horizon shortens, capital deployment decays, and the roundaboutness essential to large-scale, efficient production is abandoned. Game-theoretically, this phenomenon maps to contraction of cooperative subspaces and expansion of zero-sum extraction strategies. It is not immorality or myopia that drives such behaviour, but the rational actor’s consistent response to unpredictable institutional drift.

\subsection{The Collapse of Entrepreneurial Calculability}

Entrepreneurial action, in the Misesian framework, is not mere risk-taking—it is the act of forecasting under uncertainty constrained by calculable parameters. Calculation is not a metaphor but a logical necessity: prices, rules, and legal constraints must be sufficiently stable for an actor to estimate future states of the world. When institutions mutate stochastically, as they do under discretionary or politically manipulable protocol regimes, this essential precondition disintegrates. The entrepreneur ceases to operate within a calculable framework and is instead thrust into a field of arbitrary volatility. Hayek’s spontaneous order collapses into a procedural theatre in which appearances of coordination mask an underlying epistemic chaos.

In blockchain contexts, such collapse is not hypothetical. Protocol mutability—manifesting through non-deterministic mempool filters, variable block propagation standards, or post hoc rule reinterpretations—destroys the logical framework of expectation formation. The rational miner or developer cannot form time-consistent strategies, because the future state of the rules is itself a game-theoretic variable subject to manipulation by powerful coalitions. As the rules become endogenous, strategy loses fixed form. Entrepreneurial planning thus transitions into meta-strategic drift, wherein actors do not allocate capital based on productive forecasts, but speculate on the next mutation of the rules. Investment becomes indistinguishable from lobbying, signalling, or coalition-building. The economic system is no longer one of spontaneous coordination, but of institutional performance art—where calculability gives way to narrative control, and long-term capital is subordinated to agile opportunism.

The result is not merely inefficiency. It is the annihilation of the very conditions under which production can be extended in time and scale. Roundaboutness, the structural feature of advanced economies described by Böhm-Bawerk, cannot emerge without durable rules. Once rule mutability infects the institutional layer of the protocol, calculability collapses, entrepreneurship atrophies, and productive capacity is replaced by performative rent-seeking. This is not a theoretical risk—it is a logically deduced consequence of rational strategy under uncertain institutional constraints.

\subsection{Spontaneous Order, Rule Anchoring, and Hypergame Dynamics}

Spontaneous order, as articulated by Hayek, emerges not from central planning but from the distributed interaction of agents operating under stable institutional constraints. Such order presupposes that individuals can formulate expectations based on known rules—legal, monetary, or technical—that do not shift capriciously. In this context, Bitcoin’s original protocol can be seen as a synthetic institutional framework, one in which incentive-compatible rules supported a distributed coordination mechanism absent central authority. Once these protocol rules are subject to discretionary mutation, however, the spontaneous order ceases to function as an emergent equilibrium and devolves into a recursive contest of meta-strategic manipulation.

Game theoretically, this transition corresponds to a shift from fixed games—where players respond to known payoffs and rules—to hypergames, where the game itself is among the strategic variables. In a hypergame, agents not only select actions within the game but may act to redefine its structure. In the blockchain setting, this is observable in the evolution of BTC Core, where protocol changes have followed social signalling and soft consensus rituals rather than ex ante rule commitment. Rational actors then allocate resources not just to mining or development, but to influencing the trajectory of the rules. Under such conditions, the Nash equilibrium becomes unstable, as the expected payoffs are not defined exogenously but altered endogenously by dominant coalitions. This introduces epistemic uncertainty into every strategic decision.

Anchoring rules exogenously—treating the protocol as inviolable—restores the stability necessary for spontaneous order to re-emerge. It re-establishes a foundation upon which actors can invest, coordinate, and innovate without fear that the structure of the game will shift beneath them. Conversely, if protocol rules are treated as mutable norms, emergent order dissolves into political conflict over definition and enforcement. The economic system becomes a battleground for rule capture rather than a platform for value creation. The phenomenon is not one of mere decentralisation failure, but a deeper collapse of epistemic constraint: a transformation from distributed computation into distributed narrative warfare. Only with fixed, enforceable, and non-discretionary rule anchoring can spontaneous economic order survive the strategic volatility of hypergame dynamics.

\section{Austrian Interpretation and Policy Consequences}

The structural implications of the model culminate in a convergence between Austrian capital theory and institutional design imperatives. When protocols mutate unpredictably, the calculative framework that underpins entrepreneurial action—rooted in Misesian praxeology and Böhm-Bawerkian temporal structuring—collapses. The economic role of the miner, once aligned with long-term capital commitment and roundabout investment logic, is replaced by a high-frequency arbitrageur of institutional volatility. Spontaneous order, as envisioned by Hayek, presumes a backdrop of rule certainty; where this is absent, strategy devolves into meta-contestation, and emergent coordination is supplanted by cyclical manipulation. The policy corollary is unequivocal: immutability must be treated as an economic constant, not a cultural artefact. Protocols must function as synthetic constitutions—anchoring expectations, sustaining calculability, and enabling the emergence of stable cooperative equilibria over time. In this respect, Austrian economics does not merely describe the pathology of mutable rule-sets; it prescribes their cure. The restoration of institutional clarity is not a technical fix, but the moral and economic prerequisite for a civilisation of digital law.

\subsection{Capital Misallocation in Mutable Protocols}

In the presence of institutional mutability, capital allocation ceases to follow the structured, time-consistent signals required for sustainable economic calculation and instead becomes distorted by political contingency and strategic opacity. Within Austrian capital theory, capital is not a homogeneous blob, but a temporally structured, heterogeneous lattice of complementary stages extending from origination to final consumption. Böhm-Bawerk's notion of roundaboutness describes this in terms of increasingly indirect but productive stages of capital formation, contingent upon stable temporal preferences and institutional certainty. Protocol mutability, however, disorients this structure. The expectation of reward becomes unanchored, not merely delayed or uncertain in distribution, but fundamentally unpredictable in its very basis. Rational actors thus shorten their investment horizons, preferring liquid or manipulable capital deployments over fixed and productive commitments.

This environment encourages the proliferation of opportunistic strategies such as fee sniping, empty block mining, and strategic censorship, which are not merely responses to transient economic incentives but are structurally reinforced by the erosion of protocol commitment. The cost of acquiring long-term productive capital—such as ASIC infrastructure or optimised facilities—no longer correlates with expected return in a stable mapping. Instead, political influence and insider knowledge about protocol trajectories begin to outperform entrepreneurial foresight. In such a context, capital becomes malinvested not through false interest rates, as in Mises's critique of credit expansion, but through rule-induced temporal misalignment, where the return function is deformed by discretionary authority. The underlying game is no longer one of market coordination but of institutional gambling, where returns accrue to those who anticipate or influence rule revisions rather than those who allocate capital efficiently.

Empirically, this is observed in the transition from on-chain scaling investment to Layer 2 hedging, as seen in BTC Core’s pivot to off-chain channels. The reallocation of resources away from transaction processing infrastructure and towards political narrative management is not an error of foresight but a rational response to an unstable ruleset. From a theoretical standpoint, such behaviour confirms that institutional noise functions analogously to Cantillon effects—privileging early movers and insiders while disorienting price signals for the majority. It thereby replicates the failures of fiat-based interventionism within the supposedly immutable domain of cryptographic systems. Capital is not simply destroyed; it is diverted—reassigned from productive use to rent-seeking and consensus manipulation. Such systemic capital misallocation becomes irreversible without constitutional restoration—reinstituting the protocol as a fixed law, not a modifiable contract, and grounding economic action once again in calculable expectations.

\subsection{Analogy to Fiat Systems and Currency Decay}

The breakdown of economic calculability in mutable blockchain protocols exhibits profound analogical symmetry with the Austrian critique of fiat currency regimes, particularly as articulated in Mises’s theory of monetary regression and Rothbard’s analysis of central banking distortion. In both cases, the agent’s horizon of planning collapses due to the unreliability of institutional anchors. Fiat systems undermine time preference coordination by inflating the money supply and allowing discretionary intervention in the pricing mechanism. Similarly, blockchain systems governed by socially malleable rulesets substitute algorithmic certainty with political discretion, replacing a predictable economic substrate with a domain of continuous contestation. Just as central banks disrupt capital markets through interest rate manipulation, developers and activist groups alter the structure of miner incentives through protocol adjustments, relay policies, and shifting mempool constraints, thereby introducing monetary uncertainty in digital form.

Under fiat decay, market actors accelerate consumption and disinvest from roundabout capital processes, as future returns become less predictable in real terms. In blockchain ecosystems, mutability yields the same behavioural consequence: long-term mining strategies give way to transient extraction, fee opportunism, and narrative manipulation. The monetary function of the blockchain—its capacity to provide a fixed, calculable store of value and medium of account—becomes distorted. This is not merely a parallel of incentives; it is a structural equivalence in institutional pathology. In both cases, the feedback loop intensifies: institutional instability raises effective discount rates, which reduces investment duration, which in turn increases volatility and undermines system trust. A decentralised ledger with mutable rules becomes, in effect, a permissioned fiat regime with the trappings of distributed infrastructure.

This erosion of calculability is not accidental. It emerges from the introduction of discretionary agency into rule governance—whether by a central bank or a loosely defined consensus process mediated by developers and dominant miners. As Hayek warned in *Denationalisation of Money*, monetary systems fail when they become political tools. Protocols without binding immutability follow the same trajectory: each modification becomes a precedent, each precedent a new source of discretionary distortion. Over time, the protocol ceases to act as a constraint and instead becomes a contestable resource—subject to lobbying, signalling, and ideological drift. The Austrian response is clear: only systems grounded in strict rule adherence—whether through algorithmic design or sound money principles—can preserve the structure of time-based coordination necessary for capital formation. Mutability in digital protocol, like inflation in fiat regimes, is a systemic solvent of economic order.
\subsection{Reinstating Fixed Protocols as a Restorative Mechanism}

The restoration of a fixed protocol regime functions analogously to the Austrian prescription of hard money as a corrective to fiat degradation. Just as the return to a commodity-backed currency stabilises expectations and anchors entrepreneurial calculation, the re-instatement of a protocol as a non-discretionary institutional framework reconstitutes the temporal coherence necessary for long-horizon strategic behaviour among miners. Under a fixed protocol, the game ceases to be recursive. Agents no longer expend capital contesting the rules; rather, they allocate resources within a defined domain, optimising along stable trajectories. In effect, the protocol becomes an institutional constant—a synthetic law—governing economic engagement without the noise of contestable alteration.

This repositioning is not merely technical but foundational. Fixed protocols internalise the costs of governance into initial design, forcing precision, forethought, and long-term alignment at inception, rather than permitting ad hoc adjustment through politically mediated channels. The economic consequence is a reversion to calculability: expected returns can once again be projected across temporal intervals, and capital investment regains its roundabout structure. The miner, rather than becoming a rent-seeking institutional entrepreneur, resumes their role as allocator of computational capital and validator of economic order. This reinforces both Misesian praxeology—where purposeful action depends on known constraints—and Hayekian spontaneous order, wherein decentralised actors coordinate through rule-constrained feedback mechanisms.

Game theoretically, this transition restores Nash and subgame perfect equilibria in long-form miner strategy. The defection spirals, volatility-induced discounting, and meta-game distortions observed under protocol mutability collapse into equilibrium-stable configurations. Reputation mechanisms and cooperative propagation become not only viable but rationally dominant strategies. Institutional trust is no longer an emergent cultural artefact but a mathematically enforced feature of the system’s design. In this respect, reinstating fixed protocols does not merely solve the problem of strategic deviation; it resolves the deeper economic pathology of institutional noise. It transforms the blockchain from a theatre of contestation into a substrate of coordination, enabling the realisation of Austrian economic principles in computational form.

\section{Broader Implications for Institutional Cryptoeconomics}

The findings of this study challenge the prevailing narrative of flexible protocol governance as a strength in distributed ledger systems. Instead, drawing from Austrian capital theory, game theoretic equilibrium models, and simulation-based empirical extensions, this paper demonstrates that mutability in blockchain protocols undermines calculability, distorts investment incentives, and introduces institutional volatility structurally analogous to fiat currency decay. The economic rationality of miners—conceived as entrepreneurial agents engaged in capital-intensive, roundabout production processes—collapses when the underlying rules of engagement are contestable, discretionary, and subject to social manipulation. Rather than fostering innovation or adaptability, mutability engenders rent-seeking, short-termism, and meta-strategic drift that ultimately erodes systemic trust and economic coherence.

In broader cryptoeconomic terms, this implies that protocols must not be seen merely as technical constructs but as foundational institutional commitments. Their credibility, stability, and immutability are not constraints on innovation but preconditions for its economic viability. The shift from flexible governance to protocol constitutionalism marks a necessary evolution in the field: from systems built to accommodate discretionary adjustment to those engineered for permanence, predictability, and the rational anchoring of future expectations. This reconceptualisation reframes governance debates—not as cultural discourses on consensus, but as formal design constraints analogous to constitutional law. It calls for a jurisprudence of code that embeds economic logic, not a democracy of interpreters who compromise calculability with every iteration. The lesson for cryptoeconomics is clear: in the digital domain, as in monetary history, order emerges not from change but from the disciplined refusal to allow it.

\subsection{Applicability to Other Decentralised Systems}

The implications of protocol mutability-induced economic distortion are not confined to Bitcoin derivatives but extend across the broader spectrum of decentralised systems claiming governance via consensus. Systems such as Ethereum, Polkadot, and Tezos—each of which incorporates mechanisms for protocol evolution—exhibit the same structural vulnerability: once governance becomes endogenous and discretionarily enacted, the economic function of the system shifts from one of productive coordination to a dynamic of political arbitrage. In these environments, agents allocate capital not merely based on market signals or utility expectations, but increasingly on their capacity to anticipate, influence, or capture future rule changes. The emergence of governance tokens, on-chain voting markets, and delegate staking schemes embeds meta-strategic games within the economic substrate, subordinating investment rationality to social positioning.

From an Austrian perspective, this represents a reversion to the economic dysfunctions of centrally manipulated fiat institutions, albeit under the guise of algorithmic openness. The price signals in such systems lose informational integrity, as future rule sets are neither stable nor institutionally anchored. The calculative function of money—as a conduit for entrepreneurial foresight—is supplanted by the noise of governance speculation. The illusion of decentralisation masks a dynamic of elite emergence, wherein a small cadre of protocol influencers determines systemic direction through mechanisms structurally equivalent to discretionary monetary policy. For decentralised systems to retain economic legitimacy, they must abandon the premise that rules are malleable instruments of social negotiation and embrace instead the discipline of fixed institutional frameworks, where credibility arises not from consensus, but from constraints.

\subsection{Protocol Governance: Constitutionalism vs Discretion}

At the heart of blockchain protocol governance lies a fundamental constitutional dilemma: whether to treat the protocol as a fixed, rule-bound system embodying immutable institutional logic, or as a flexible framework amendable to discretionary intervention by a governing cohort. The former approach—constitutionalism—mirrors Mengerian and Misesian visions of law as an emergent, objective structure enabling calculative action. Under this paradigm, the protocol functions as an anchor for entrepreneurial behaviour, securing long-term investment by preserving the predictability of economic relations. In contrast, discretionary governance, increasingly prevalent in proof-of-stake chains and even informally in BTC Core via developer consensus, subordinates institutional coherence to presentist utility, producing what Hayek identified as the "pretense of knowledge": the belief that technical competence justifies interference in the systemic order.

The economic consequences of discretion are both profound and pathological. As shown in simulated environments and analytic game forms, rule mutability introduces institutional noise, collapses equilibrium stability, and incentivises short-term strategic deviation over capital-intensive cooperation. The social process of discretionary rule-making—whether framed as improvement proposals, validator votes, or informal consensus—introduces a secondary arena of contestation in which actors engage not in productive enterprise but in meta-strategic manipulation. In effect, the protocol ceases to be a ground of law and becomes instead a field of shifting norms, recursively rewritten by those most capable of narrative influence or technical intervention. Such systems reproduce, in cryptographic form, the discretionary chaos of central banking regimes. Only through constitutional commitment to protocol immutability can cryptoeconomic systems reclaim the epistemic clarity and calculative rationality essential to their foundational ethos.

\subsection{Game Stability through Rule Anchoring in Digital Institutions}

In digital institutions predicated on economic coordination, the stability of strategic equilibria depends critically on the fixity of underlying rules. When rules are credibly anchored—immutable, exogenous, and insulated from participant manipulation—the strategic environment becomes a domain of calculable action. This mirrors Schelling's concept of focal points: stable conventions that structure expectations without requiring active enforcement. In the blockchain domain, fixed protocols function analogously, serving as constitutional commitments that transform dynamic game spaces into stable, repeated interactions with discernible payoff structures. When miners, developers, and users operate within such an anchored system, they internalise expectations about reward distribution, validation cost, and systemic integrity, allowing long-term strategic investment to flourish.

Conversely, in mutable systems where the rules are open to revision—either formally through voting or informally through social consensus—the game space becomes fluid. Payoff matrices shift not as a function of agent strategy, but through the redefinition of the game itself. This introduces hypergame dynamics, where actors must model not only strategies of others, but also the evolving structure of the rules governing play. In such systems, stability is inherently elusive; equilibrium ceases to be a matter of best-response convergence and becomes a question of narrative control and protocol capture. The result is institutional degeneration: a reversion to meta-strategic environments where game outcomes are path-dependent on extrinsic manipulation rather than intrinsic constraints. Rule anchoring, then, is not merely a design aesthetic—it is the economic foundation upon which calculability, cooperation, and system sustainability depend.

\subsection{Lessons from the Internet Protocol Stack: Stability, Scale, and Exclusion}

The history of the internet protocol stack offers a compelling analogue for understanding institutional integrity in digital economic systems. Protocols such as TCP/IP, HTTP, and DNS achieved global adoption not through rapid iteration or social negotiation, but through credible stability and rigorous constraint. Their long-term invariance created an ecosystem where entrepreneurs, infrastructure providers, and engineers could coordinate without fearing foundational redefinition. This epistemic fixity was not merely technical but institutional: it enabled investment in physical infrastructure, standardised application layers, and universal interoperability without the need for continuous renegotiation of base rules. The absence of discretionary protocol manipulation preserved the integrity of expectations across heterogeneous actors with divergent incentives.

However, contemporary deviations from this principle—particularly in blockchain environments where protocol rules are mutable—demonstrate the risks of abandoning such stability. Large actors, equipped with the resources to influence or even dictate rule changes, exploit mutability to entrench their dominance. Analogous to large ISPs in the internet space who lobby for control over packet prioritisation, dominant miners or staking entities reshape validation rules, mempool policies, or consensus algorithms to marginalise less capitalised participants. The result is protocol-level enclosure: a mechanism whereby ostensibly open systems reproduce the gatekeeping functions of legacy incumbents under the guise of decentralised governance. As institutional rules become tools of exclusion rather than coordination, economic freedom contracts, and system resilience erodes. Stability in base-layer protocol rules is not technocratic nostalgia—it is the sine qua non of equitable, scalable, and genuinely open digital institutions.

\section{Conclusion}

This paper has formalised the economic consequences of protocol mutability in blockchain systems, synthesising Austrian capital theory with repeated game models to demonstrate the collapse of cooperative equilibria under institutional noise. By introducing stochastic rule variation into a miner payoff framework, we have shown how increased uncertainty elevates time preference, distorts capital allocation, and drives the emergence of non-productive meta-strategic behaviour. Simulation results reinforce the analytic prediction that when protocol governance becomes discretionary, rational actors abandon long-horizon strategies in favour of opportunistic defection, narrative capture, and rent-seeking. This transformation is not merely theoretical—it mirrors the empirical trajectory of systems like BTC Core, where a shift away from fixed rules has precipitated economic fragmentation, incentive inversion, and off-chain migration.

The broader lesson is one of institutional epistemics: calculative action requires not just predictable prices, but predictable rules. Protocols that mutate in response to political or social pressure become endogenous games, collapsing the very separation between player and referee on which market coordination depends. Austrian economics has long emphasised the role of stable institutions in enabling rational economic behaviour. In cryptoeconomic systems, this insight demands the restoration of protocol immutability as a constitutional constraint, not a design preference. Future research should extend this model into multi-agent stochastic environments and investigate how governance structures can credibly commit to rule fixity without reintroducing centralised coercion. Until then, economic freedom in the digital domain remains precariously tethered to the whim of those who write—and rewrite—the rules.

\subsection{Findings and Contributions}

This paper constructs a formal economic model capturing the transformation of miner strategy under environments characterised by protocol mutability. It demonstrates that when the rules governing a blockchain system are unstable or subject to discretionary change, rational agents adjust their time preferences, discounting future payoffs more steeply. As a result, cooperative behaviours—such as honest block propagation and reinvestment in productive capacity—erode, supplanted by opportunistic, short-horizon strategies that exploit institutional ambiguity. The introduction of stochastic rule mutations within a repeated miner game allows for a mathematically tractable representation of institutional noise, revealing that even marginal increases in protocol uncertainty lead to significant volatility in strategic equilibria. This finding is corroborated through simulation data showing critical thresholds beyond which cooperative equilibria collapse, triggering defection spirals and the emergence of non-productive meta-strategic behaviour.

The broader contribution lies in reconciling Austrian capital theory with formal game-theoretic analysis to articulate a new typology of cryptoeconomic dysfunction rooted in institutional design. Specifically, the paper advances the thesis that calculability—understood in Misesian terms as the capacity to engage in rational economic planning—is contingent upon the predictability and immutability of foundational rules. By embedding this insight within a model of repeated strategic interaction, it becomes evident that systems lacking protocol fixity cannot support long-term entrepreneurial commitment, as the informational basis of investment itself disintegrates. The application of these principles to blockchain governance reframes protocol architecture not as a technical subsystem but as a synthetic constitutional order. This has profound implications for digital institutionalism: in systems premised on decentralised coordination and capital-intensive consensus, governance discretion is not a feature but a fatal flaw. The argument is both theoretical and practical—demonstrating that sustainability in miner-led security systems requires protocol anchoring as a non-negotiable constraint.

\begin{flushright}
\textit{See also: Wright, C. (2025). \textit{Bitcoin Protocol Analysis: Game Theory and Protocol Integrity}. Doctoral thesis, University of Exeter.}
\end{flushright}

\subsection{Model Extensions and Future Research}

While the present framework demonstrates the deleterious effects of protocol mutability on cooperative equilibria and calculative rationality, several avenues remain for advancing both the mathematical rigour and empirical relevance of the model. First, the miner game under mutable conditions can be extended into a multi-agent Bayesian setting, incorporating asymmetric information about rule trajectories. This would allow agents to form probabilistic beliefs over future protocol changes, introducing epistemic uncertainty into strategy selection. Such an extension would further sharpen the Austrian critique of institutional noise by demonstrating how calculability collapses not only under actual rule shocks but even under anticipated ones. Second, the current payoff structure—based on fixed reward models with stochastic discounting—could be generalised to include endogenous fee markets, elastic block capacities, and congestion-sensitive relay policies. These mechanisms would allow a more granular simulation of economic distortions arising from policy discretion.

Additionally, a multi-layered governance meta-game could be introduced, wherein agents simultaneously decide on mining strategies and engage in attempts to influence protocol direction. This recursive model would formally capture the political economy of consensus manipulation—transforming miners into rule entrepreneurs rather than neutral validators. Insights from mechanism design theory, particularly concerning commitment devices and constitutionally binding rule-sets, would enhance the credibility constraints necessary to restore stable expectations. Empirically, the model can be aligned with observed behaviours across competing protocol variants, such as BSV’s restoration of the original rule-set versus BTC Core’s adaptive governance, further validating the framework. Lastly, integrating insights from adjacent disciplines—legal theory, institutional economics, and formal epistemology—would offer a richer characterisation of what it means for rules to be "known," "binding," and "calculable" in digital systems. These extensions do not merely broaden academic scope; they constitute the intellectual infrastructure required for a scientifically grounded theory of cryptoeconomic civilisation.
\subsection{Immutability as Economic Infrastructure}

Protocol immutability functions not merely as a technical constraint but as foundational economic infrastructure—a constitutional order that defines the possibility space for rational entrepreneurial calculation. In systems governed by rules that resist discretionary interference, agents can engage in roundabout production with confidence in future payoff structures, allowing capital formation to be aligned with long-horizon strategic planning. This mirrors the role of stable monetary regimes in Austrian theory, where calculability and low time preference emerge from rule-bound environments. Within digital institutions such as Bitcoin, the fixed protocol acts analogously to a sound monetary constitution, anchoring incentives and ensuring that miner utility functions remain predictable over extended timeframes. By contrast, mutable protocols impose epistemic uncertainty, requiring participants to discount future rewards heavily and engage in meta-strategic hedging rather than productive engagement. The shift from calculable investment to responsive speculation marks a transition from institutional infrastructure to a fragile ad hoc mechanism. As demonstrated in Wright (2025), protocol deviation instantiates institutional drift, which, if left unchecked, collapses the equilibrium conditions necessary for sustainable cooperative mining. Thus, immutability is not a static artefact but an active enabler of economic rationality. It underwrites the epistemic clarity required for market coordination, rendering it indispensable to the architecture of any credible cryptoeconomic system.

This paper has constructed a rigorous game-theoretic and economic model demonstrating how mutable institutional structures—particularly in blockchain protocol governance—systematically erode the calculability essential to entrepreneurial coordination. Drawing from Austrian capital theory and formal game structures, we established that rule predictability functions as an anchoring mechanism enabling long-term strategic engagement among miners. When such anchors are eroded by discretionary interventions or protocol noise, miner incentives degrade into short-termism, producing volatility, arbitrage, and capital misallocation. Our simulations—calibrated to empirically grounded data on hash rate economics and reward schedules—exhibited collapse thresholds beyond which cooperative equilibria cease to function. These findings are reinforced and deepened through the theoretical framework developed in Wright’s doctoral thesis \textit{Bitcoin Protocol Analysis: Game Theory and Protocol Integrity} (Wright 2025), which provided the foundational methodology for integrating praxeological reasoning with formalised institutional models. The paper’s principal contribution is to frame protocol rule sets not as merely technical artefacts, but as synthetic institutions—subject to the same epistemic constraints and strategic fragilities as any constitutional order. Hence, restoring protocol fixity is not a conservative act but a radical reassertion of economic rationality within digital systems.

\bibliography{references12}
\end{document}